\documentclass[12pt]{article}

\usepackage{dcolumn}
\usepackage{bm}
\usepackage[latin1]{inputenc}
\usepackage[spanish,english]{babel}
\usepackage{amsfonts}
\usepackage{amssymb}
\usepackage{graphicx}
 \usepackage{setspace}
 \usepackage{verbatim} 

\newcommand{\be}{\begin{equation}}
\newcommand{\ee}{\end{equation}}
\newcommand{\bea}{\begin{eqnarray}}
\newcommand{\eea}{\end{eqnarray}}

\begin{document}

\begin{titlepage}

\begin{center}
{\large \bf Stimulated creation of quanta during inflation and the observable universe \footnote{Awarded with the First Prize in the Gravity Research Foundation Essay Competition 2011} } \\
\end{center}

\begin{center}
Ivan Agullo\footnote{ agullo@gravity.psu.edu}\\ 
\footnotesize \noindent {\it Institute for Gravitation and the Cosmos, Physics Department,
Penn State, University Park, PA 16802-6300, USA.}\\
\end{center}

\begin{center} Leonard Parker\footnote{leonard@uwm.edu}\\

\footnotesize \noindent {\it Physics Department, University of
Wisconsin-Milwaukee, P.O. Box 413, Milwaukee, WI 53201, USA}

\end{center}
\begin{center}
{\it March 21, 2011}
\end{center}

\begin{abstract}
Inflation provides a natural mechanism to account for the origin of cosmic structures. 
The generation of primordial inhomogeneities during inflation can be understood via the spontaneous creation of quanta from the vacuum. We show that when the corresponding {\it stimulated} creation of quanta is considered, the characteristics of the state of the universe at the onset of inflation are not diluted by the inflationary expansion and can be imprinted in the spectrum of primordial inhomogeneities.  The non-gaussianities (particularly in the so-called squeezed configuration) in the cosmic microwave background and galaxy distribution can then tell us about the state of the universe that existed at the time when quantum field theory in curved spacetime first emerged as a plausible effective theory.

\end{abstract}

\end{titlepage}

One of the deepest problems of current research concerns the formulation and experimental confirmation of a comprehensive theory from which our present knowledge of  gravitation and particle physics will emerge.  The progress of physics has long been guided by experimental and observational results; and these are very difficult to come by at the energy scales of such a comprehensive theory.  

Our gap in knowledge has partly been illuminated by the quantum theory of fields in curved spacetime (QFT in CST). In this framework, one studies quantum field theory in spacetimes described by classical metrics, as in general relativity, in a regime where we are confident about the validity of both theories. One of the most remarkable physical outcomes of QFT in CST is the phenomenon of gravitationally-induced spontaneous creation of quanta in curved spacetimes, as first pointed out and analyzed by one of us \cite{parkerthesis, parker69, parker68-71} in the cosmological context of an expanding universe.   

In the following two decades, fundamental implications of this phenomenon were obtained. Hawking \cite{hawk1} showed that black holes produced by gravitational collapse would emit thermal radiation.  In the cosmological context, when the idea of inflation was investigated \cite{inflation},  it was proposed  \cite{inflation2} that the inflationary expansion of the universe leads to a spectrum of perturbations that can seed the cosmic inhomogeneities we observe today. 

Black hole thermal radiance has not been detected yet. However, the high degree of precision in measurements of the temperature fluctuations of the cosmic microwave background (CMB) and the large scale structure (LSS) achieved during the last decade, together with the new generation of observational missions, has opened a fascinating window to measure the predictions of QFT in CST.

The computation of the spectrum of non-uniformities created during inflation requires a basic assumption regarding the quantum state describing scalar and tensor metric perturbations at the onset of inflation. Such a state, in the context of slow-roll inflation, is chosen as a natural extension of the Bunch-Davies vacuum \cite{bunch-davies}. The election of the vacuum can seem unnatural due to our ignorance of the physics in the early universe before inflation. It has been argued that the exponential expansion during inflation will dilute any possible quanta present in the initial state and will drive any arbitrary state to the vacuum. That justifies the vacuum state as the most natural choice. However, as first pointed out in \cite{parkerthesis,parker69} (see Eq. (53) in \cite{parker69}), the gravitationally-induced spontaneous creation of quanta in a general  expanding universe is accompanied by the corresponding stimulated process if there are quanta already present in the initial state. The stimulated creation process during inflation is governed by the {\em same} quantum amplification factor as is responsible for the spontaneous creation of quanta of the perturbation field from the vacuum. This amplification factor grows enormously during inflation and compensates for the dilution of the quanta initially present, keeping their average number density constant during inflation. The objective of this essay is to study the effects in the observable universe of the stimulated creation of quanta when a general initial state is considered. Our detailed results are given in \cite{agulloparker}.

We focus our computation in scalar perturbations in single-field inflation.  Scalar perturbations are conveniently characterized by the gauge-invariant comoving curvature perturbations ${\cal{R}}(\vec{x},t)$ (see, for instance, \cite{Weinberg2008}).  The most general initial state for scalar perturbations can be described as a mixed state characterized by a statistical density operator $\rho$. Such a mixed state is natural within the context of inflation in which the observable universe results from the enormous expansion of a small patch of a much larger universe. The state of such a patch would naturally be a mixed state even if the global state of the larger universe were a pure state because many features of the global state would not be accessible to our observable universe \cite{vonneumann}.

We consider an isotropic density operator $\rho$ formed from a mixture of pure states having definite numbers of initial quanta (for the explicit form, see \cite{agulloparker}). A more general density operator would leave our main conclusions unchanged. The extension to anisotropic initial states as a potential source for anisotropies and other related anomalies that may be present in the CMB and LSS would be straightforward. As shown in \cite{lyth,boyanovsky,holman}, by imposing renormalizability of $\rho$ and negligible back reaction on the inflationary expansion, the average number of initial quanta, ${\rm Tr}[\rho N_{\vec{k} }]$, in a mode with coordinate momentum ${\vec k}$, is constrained to be not much larger than 1.

To leading order, the spectrum of perturbations created in the mixed state $\rho$ during inflation is characterized by the two-point function in momentum space or, equivalently, the power spectrum $P^{\rho}_{\cal{R}}(k)$, where $k\equiv |\vec{k}|$.
A straightforward computation gives the following expression (see \cite{agulloparker} for details)
\be  \label{newpowspect} P^{\rho}_{{\cal R}}(k)= P^0_{{\cal R}}(k) (1+2 \, {\rm Tr}[\rho N_{\vec{k} }]) \ ,\ee
where $P^0_{{\cal R}}(k)$ is the power spectrum resulting from a pure vacuum state at the onset of inflation.

The observation \cite{wmap7} of the amplitude of the power spectrum and spectral index, $n_s$, characterizing its dependence on $k$ can be used to constrain the average number of initial quanta ${\rm Tr}[\rho N_{\vec{k} }]$.  However, the power spectrum alone has limited potential in revealing detailed information about the initial state, as it only constraints the quantity $\left| \frac{d\ln{ (1+2{\rm Tr}[\rho N_{\vec{k} }])}}{d \ln k}\right| $ to be not much bigger than the slow-roll parameters $\epsilon \sim |\delta|\sim 1/100$, defined in terms of the Hubble rate, $H(t)$, as $\epsilon \equiv -\dot{H}/H^2$ and $\delta \equiv \ddot{H}/(2\dot{H} H)$.

On the other hand, it has been shown (see, for instance, \cite{whitepaper}) that non-gaussianities in the distribution of the perturbations give a sharp tool to probe many aspects of inflation that are not uniquely determined by their power spectrum alone. These non-gaussianities are characterized by their three-point function in momentum space, or equivalently the bispectrum $B_{{\cal{R}}}(\vec{k}_1,\vec{k}_2,\vec{k}_3) $.
We compute the bispectrum to leading order in the perturbations, when their initial state is given by a density operator $\rho$, by generalizing the pioneering computation by Maldacena  \cite{maldacena}, where the vacuum state was considered. The complete expression for our calculated  bispectrum can be found in  \cite{agulloparker}. The presence of initial perturbations introduces new interesting features. Our results show that when particles are present in the initial state, there are new contributions to $B^{\rho}_{{\cal{R}}}$ that come from perturbative interactions among the quanta produced by stimulated creation resulting from the presence of initial particles. As a consequence, we find an enhancement of $B^{\rho}_{{\cal{R}}}$ for certain configurations of the momenta $\vec{k}_1$, $\vec{k}_2$, and  $\vec{k}_3$. Remarkably, the largest enhancement appears  for configurations in which two of the momenta are much bigger than the third one, for instance $k_1\approx k_2\gg k_3$, the so-called squeezed configuration. The bispectrum in the squeezed configuration is generally parametrized as \cite{komatsuspergel}
\be  B_{{\cal{R}}}(\vec{k}_1,\vec{k}_2,\vec{k}_3) = \  P_{\cal{R}}(k_1) P_{\cal{R}}(k_3) \frac{12}{5} f_{NL}\ . \ee
In the case of single-field inflation with no particles in the initial state we obtain $f_{NL}^0=\frac{5}{12} (1-n_s^0)=\frac{5}{12}(4 \epsilon+2\delta)$, in agreement with \cite{maldacena}. Therefore, for the vacuum state $f_{NL}^0\approx {\mathcal O}(\epsilon,\delta)\ll1$ for single-field slow-roll inflation.

In the case when the density operator $\rho$ involves initial quanta with average numbers ${\rm Tr}[\rho N_{\vec{k} _i}]$ of order or greater than 1, we find
\be \label{bis} f_{NL}^{\rho} \approx  \frac{5}{3} \epsilon \frac{k_1}{k_3}  \left( \frac{ 2 \,  {\rm Tr}[\rho N_{\vec{k} _1} N_{\vec{k} _2}]+{\rm Tr}[\rho N_{\vec{k} _1}]+{\rm Tr}[\rho N_{\vec{k} _2}]}{{\rm Tr}[\rho (2 
N_{\vec{k} _1}+1)]{\rm Tr}[\rho (2 N_{\vec{k} _2}+1) ]} \right) \ . \ee
In obtaining (\ref{bis}), we have taken into account that $\vec{k}_3$ cannot be arbitrarily small. This is because $\vec{k}_1$, $\vec{k}_2$ and $\vec{k}_3$ are modes that we want to observe in the CMB and LSS, so their wavelengths today cannot be much larger than the observable universe. If there are no significant correlations between initial particles with different momenta,
the factor in parenthesis in (\ref{bis}) takes the value $\sim 1/2$, independently of the exact values of the ${\rm Tr}[\rho N_{\vec{k} _i}]$. Then $f_{NL}^{\rho} \approx \frac{5}{6} \epsilon \frac{k_1}{k_3}$, and the relative enhancement of the non-gaussianities by the presence of initial quanta is given by
\be \label{ratio} \frac{f_{NL}^{\rho}}{f_{NL}^{0}}\approx \frac{k_1}{k_3} \frac{2 \epsilon}{4\epsilon +2 \delta}\ . \ee

Because in the squeezed configuration we have $k_1\gg k_3$, it follows that  ${f_{NL}^{\rho}}/{f_{NL}^{0}}\gg 1$.
The range of $k$'s for which we have some confidence about the measurements of the temperature fluctuations of the CMB (i.e., for which uncertainities coming from cosmic variance, Sunyaev-Zeldovich effect, etc., can be neglected) is approximately two orders of magnitude. Therefore, the ratio $\frac{k_1}{k_3}$, and hence the relative enhancement in $f_{NL}$, can be as large as ${\mathcal{O}}(100)$. Finally, we note that in the case that the average number of initial quanta ${\rm Tr}[\rho N_{\vec{k} _i}]$ is considerably smaller than 1, additional terms not explicitly shown in (\ref{bis}) become relevant  and the vacuum result is smoothly recovered.

Non-vacuum states have been considered in previous works \cite{gasperiniveneziano,holman}. However, the large enhancement of non-gaussianities in the squeezed momentum configuration has not been considered.

The forthcoming observations of the CMB by Planck and the LSS by a variety of galaxy surveys will strongly constrain aspects (including amplitude, scale dependence, and sign) of the non-gaussianities produced during inflation. Recent studies show that the contribution to the non-gaussianities relevant in the squeezed limit ($k_1\approx k_2 \gg k_3$) will be particularly well constrained by LSS observations \cite{shandera}. The momentum dependence of $f_{NL}^{\rho}$, given by the ratio $k_1/k_3$, will offer a clear observational signature for identifying the effects of scalar perturbations present at the onset of inflation.

Our results indicate that if non-gaussianities are observed in the squeezed momentum configuration, then single-field inflation is not necessarily ruled out, as widely  accepted \cite{creminellizaldarriaga}. Instead, such an observation could be interpreted as a consequence of a non-vacuum initial state. This will allow us to learn  
about the state of the universe that existed at the time when QFT in CST first emerged as a plausible effective theory, and in turn may give valuable information about a more comprehensive theory beyond the range of validity of QFT in CST. 

\noindent { \bf Acknowledgements.} 
This work has been supported by the NSF grant PHY-0854743 and Eberly research funds of Penn State University, by the University of Wisconsin-Milwaukee RGI funds and NSF grant PHY-0503366, and by the Spanish grant FIS2008-06078-C03-02 and EXPLORA funds FIS2010-09399-E. The authors thank A. Ashtekar, J. Navarro-Salas, W. Nelson, and G. J. Olmo for many stimulating discussions.

\end{document}